\documentclass[runningheads]{llncs}
\usepackage{graphicx}
\usepackage{amssymb}
\usepackage{mathtools}
\usepackage[ruled,linesnumbered]{algorithm2e}
\usepackage{multirow}
\usepackage{siunitx}
\usepackage{booktabs}
\usepackage{makecell}
\usepackage{floatrow}
\usepackage[misc,geometry]{ifsym}
\usepackage{hyperref}
\newfloatcommand{capbtabbox}{table}[][\FBwidth]
\RestyleAlgo{ruled}
\begin{document}
\title{O2CTA: Introducing Annotations from OCT to CCTA in Coronary Plaque Analysis}
\titlerunning{O2CTA: Introducing Annotations from OCT to CCTA}
%
\author{
Jun Li\inst{1} \and
Kexin Li\inst{3,4} \and
Yafeng Zhou\inst{3,4} \and
S. Kevin Zhou\inst{2,1}{\href{mailto:s.kevin.zhou@gmail.com}{\textsuperscript{\Letter}}}
}

%
\authorrunning{J. Li et al.}
%
\institute{
Key Lab of Intelligent Information Processing of Chinese Academy of Sciences (CAS), Institute of Computing Technology, CAS, Beijing 100190, China \and
School of Biomedical Engineering \& Suzhou Institute for Advanced Research, Center for Medical Imaging, Robotics, and Analytic Computing \& Learning (MIRACLE), University of Science and Technology of China, Suzhou 215123, China \and
Department of Cardiology, Dushu Lake Hospital Affiliated to Soochow University, Suzhou 215000, China \and
Institution for Hypertension of Soochow University, Suzhou 215000, China
}
%
\maketitle              
\begin{abstract}
Targeted diagnosis and treatment plans for patients with coronary artery disease vary according to atherosclerotic plaque component. 
Coronary CT angiography (CCTA) is widely used for artery imaging and determining the stenosis degree. However, the limited spatial resolution and susceptibility to artifacts fail CCTA in obtaining lumen morphological characteristics and plaque composition. It can be settled by invasive optical coherence tomography (OCT) without much trouble for physicians, but bringing higher costs and potential risks to patients. Therefore, it is clinically critical to introduce annotations of plaque tissue and lumen characteristics from OCT to paired CCTA scans, denoted as \textbf{the O2CTA problem} in this paper. We propose a method to handle the O2CTA problem. CCTA scans are first reconstructed into multi-planar reformatted (MPR) images, which agree with OCT images in term of semantic contents. The artery segment in OCT, which is manually labelled, is then spatially aligned with the entire artery in MPR images via the proposed alignment strategy.
Finally, a classification model involving a 3D CNN and a Transformer, is learned to extract local features and capture dependence along arteries.
Experiments on 55 paired OCT and CCTA we curate demonstrate that it is feasible to classify the CCTA based on the OCT labels, with an accuracy of 86.2\%, while the manual readings of OCT and CCTA vary significantly, with a Kappa coefficient of 0.113. We will make our source codes, models, data, and results publicly available to benefit the research community.

\keywords{OCT \and CCTA \and Coronary plaque analysis.}
\end{abstract}
\section{Introduction} \label{sec.intro} 


Coronary artery disease (CAD) has become one of the leading causes of death worldwide~\cite{mozaffarian2016heart}. It results from a plaque buildup in coronary artery walls, narrowing the coronary arteries and obstructing blood delivered to the heart. A plaque comprises various substances and can be characterized as fibrous, calcified, and lipid-rich. A calcified plaque is considered stable, while non-calcified and mixed plaques are unstable and rupture-prone. Therefore, it is crucial to differentiate and evaluate coronary plaque to make appropriate patient diagnoses and treatment strategies ~\cite{cassar2009chronic}.
 
Coronary computed tomographic angiography (CCTA) has a good feasibility and reproducibility in imaging. It provides three-dimensional visualization of the coronary lumen, stenoses, and plaque features, non-invasively~\cite{maurovich2014comprehensive}. In clinical practice, the distinguishing features of CCTA are non-invasive and relatively low-cost. But the limited spatial resolution and susceptibility to artifacts hinder CCTA from throwing more punches in plaque assessment. The tissue characteristics of plaques are difficult to determine, and ruptured plaques may be misdiagnosed solely based on CCTA~\cite{yang2018coronary}.

Optical coherence tomography (OCT) exploits near-infrared light for intravascular examination and creates a detailed tissue image with extraordinarily high-resolution, cross-sectional tomographic images~\cite{fujii2015accuracy}. OCT has the advantage of differentiating tissue characteristics and provides a more detailed morphology of coronary plaques compared with CCTA~\cite{nakazato2015atherosclerotic}. Hence, OCT is considered as the gold standard for accurately determining high-risk plaque features, such as thin cap fibroatheroma. However, OCT is costly and invasive, which induces potential risks to patients.

Deep learning techniques have recently revolutionized and transformed the field of medical imaging processing~\cite{zhou2017deep,shen2017deep} and profoundly influenced automated coronary analysis~\cite{litjens2019state}. Deep learning models based on convolutional neural networks (CNNs), recurrent neural networks (RNNs), or vision transformerss~\cite{dosovitskiy2020image,Li2022TransformingMI} assist clinicians in diagnosing CAD. They grow mature rapidly in coronary stenosis detection~\cite{zreik2018recurrent,tejero2019texture,ma2021transformer}, centerline extraction~\cite{wolterink2019coronary,gao2021joint,yang2020cpr}, vessel segmentation~\cite{gao2021joint,gharleghi2022automated}, fractional flow reserve (FFR) prediction~\cite{itu2016machine,zreik2019deep}, and plaque classification~\cite{zreik2018recurrent,denzinger2019coronary} based on CCTA images; plaques classification~\cite{abdolmanafi2017deep,lee2020segmentation}, and lumen segmentation~\cite{yong2017linear,carpenter2022automated} on OCT. 

Specifically, deep learning methods are proposed to classify plaque into calcified, partially calcified, and non-calcified on CCTA~\cite{zreik2018recurrent,denzinger2019coronary}. As stated above, even experienced exports cannot identify the tissue characteristics of plaques accurately grounded on the CT attenuation pattern. It is challenging to obtain reliable annotations on plaque composition, and we cannot construct a deep learning method on just CCTA images. But it is a different situation when it comes to OCT images. 
Then it is natural to inquire: \textit{Is it possible to build reliable models to classify plaque characteristics and composition on CCTA images, with accurate and solid annotations acquired from OCT images?} We denote it as \textbf{the O2CTA problem}. If the answer is yes, CCTA will expand its boundaries on the plaque classification, and get compelling results as OCT. It is clinically meaningful to obtain finer analytical results, cost-effectively and non-invasively.

\begin{figure}[t!]
\includegraphics[width=\textwidth]{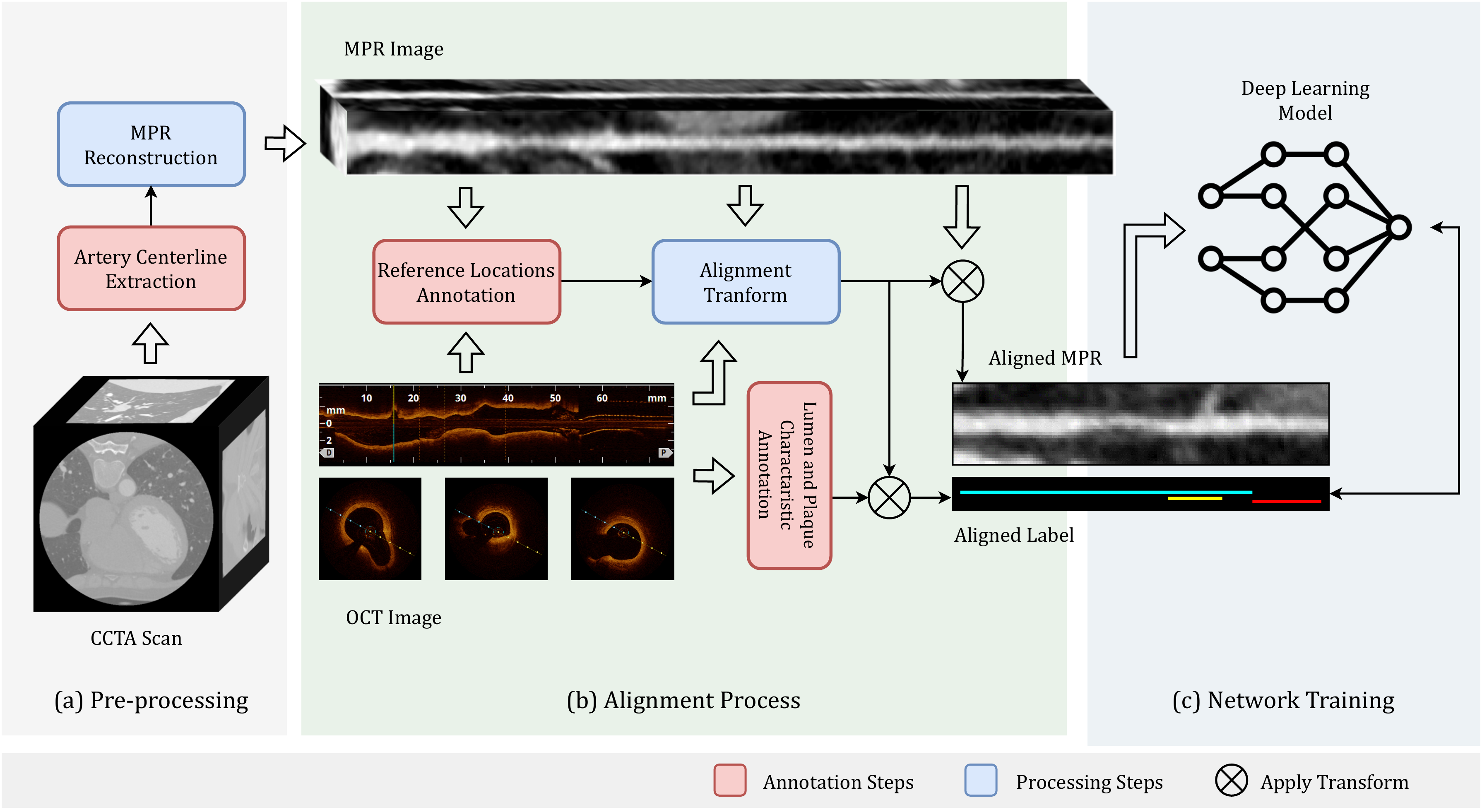}
\caption{The proposed workflow for solving the O2CTA problem includes pre-processing, alignment, and network training stages.}
\label{fig.procedures}
\end{figure}

This paper proposes the corresponding measures and processes to resolve this brand-new task, O2CTA. Fig.~\ref{fig.procedures} introduces the workflow of the proposed measure. The CCTA scans are reconstructed into multi-planar reformatted (MPR)~\footnote{Strictly speaking, this is not an MPR; rather it is an MPR using the `distorted' volume with the artery centerlines straightened.} images with the help of extracted artery centerlines. MPR and OCT images are consistent in visualization style, but they are still staggered in space. The spatial alignment requires matching locations as a reference, which need to be identified manually. The intervals of reference locations in MPR and OCT images support constructing an alignment transform, generating the aligned MPR segment, and introducing finer label to MPR space. Then, datasets are prepared to train a deep learning model and make predictions. Experimental results illustrate that the proposed method gives a promising answer to the O2CTA problem.

\section{Methodology} \label{sec.method}






\subsection{Task Formulation}
The required data for this task is retrospectively collected in a single site between 2021 and 2022, including 55 CCTA scans $\mathbf{D}_{CCTA}$ and OCT images $\mathbf{D}_{OCT}$. Each patient has a CCTA volume $X^i_{CCTA} \in \mathbf{D}_{CCTA}$ and a paired OCT image $X^i_{OCT} \in \mathbf{D}_{OCT}$ acquired within six months. 

According to coronary morphological characteristics and compositions of plaque, the $\mathbf{D}_{OCT}$ is fine-annotated into $\mathbf{Y}_{OCT}$ by an expert with five-year experience. Under the imaging quality of CCTA, we cannot generate annotations $\mathbf{Y}_{CCTA}$ with the same precision and accuracy as $\mathbf{Y}_{OCT}$. Therefore, an alignment function $\mathcal{A}$ is required to align $\mathbf{Y}_{OCT}$ into paired CCTA space:
\begin{equation}
    \mathcal{A}: \mathbf{D}_{OCT} \rightarrow  \mathbf{D}_{CCTA},\quad {\rm then} \quad \mathcal{A}: \mathbf{Y}_{OCT} \rightarrow  \mathbf{Y}_{CCTA}
    \label{eq.align}
\end{equation}
The alignment function $\mathcal{A}$ describes spatial relation between paired $X^i_{OCT}$ and $X^i_{CCTA}$. It is also termed as an alignment transform over annotations $Y^i_{OCT}$ and $Y^i_{CCTA}$. After applying $\mathcal{A}$ on $\mathbf{Y}_{OCT}$, the corresponding annotation of CCTA $\mathbf{Y}_{CCTA}$ is obtained and ready for training a deep learning model $\mathcal{F}$:
\begin{equation}
    \mathcal{F}: \mathbf{D}_{CCTA} \rightarrow \mathbf{Y}_{CCTA}
    \label{eq.model}
\end{equation}
In general, we aim to build a deep learning model $\mathcal{F}$ on $\mathbf{D}_{CCTA}$ and predict more precise and accurate results under the guidance of $\mathbf{Y}_{OCT}$. As stated above, the core issues of the solution can be degraded as spatial mapping function $\mathcal{A}$ and deep learning model $\mathcal{F}$. We detail our solutions to these two problems below.

\subsection{Alignment Function}
\label{sec.align}
The alignment function $\mathcal{A}$ is crucial in preparing annotations for the deep learning model $\mathcal{F}$. Before detailing the method $\mathcal{A}$, we need to introduce some key ingredients.

\subsubsection{Alignment Objects}
OCT presents the arterial lumen cross-sectionally, while CCTA scans visualize the entire chest and heart by 3D volumes with coronary arteries enhanced. The first issue that arises is keeping the presentation manner consistent. 

In each CCTA image $X^i_{CCTA}$, the centerline $P^i \in \mathbf{P}$ of the corresponding coronary artery in $X^i_{OCT}$ is extracted using the method previously proposed ~\cite{wolterink2019coronary}. A single seed point is placed manually, and the centerline of the interested artery is extracted between the ostium and the most distal point in $X^i_{CCTA}$. MPR image $X^i_{MPR} \in \mathbf{D}_{MPR}$ are reconstructed using extracted $P^i$, which visualizes coronary arteries in line with OCT. At this point, the alignment objects, $X^i_{MPR}$ and $X^i_{OCT}$, are ready, and Eq.~(\ref{eq.align}) is updated as:
\begin{equation}
    \mathcal{A}: \mathbf{D}_{OCT} \rightarrow  \mathbf{D}_{MPR},\quad {\rm then} \quad \mathcal{A}: \mathbf{Y}_{OCT} \rightarrow  \mathbf{Y}_{MPR}
    \label{eq.align_update}    
\end{equation}

\subsubsection{Reference Locations for Alignment}
However, $\mathbf{D}_{MPR}$ present entire arteries from the ostium to the distal, not the start- and end-points of the part of artery segments affected by plaque in $\mathbf{D}_{OCT}$. The alignment $\mathcal{A}$ still lacks spatial matching information to execute. Therefore, reference points $\mathbf{R}$ are manually screened based on locations like artery bifurcations, severe calcification, stent, and stenosis. 
\begin{equation}
    R^i = \{(r_{OCT}^j, r_{MPR}^j)\ |\ j \in [1, 2, \dots]\}
\label{eq.refer}
\end{equation}
where each case has at least one reference points pair. $r_{OCT}^j$ and $r_{MPR}^j$ refer to the exact anatomical location in $X^i_{MPR}$ and $X^i_{OCT}$, respectively. Notice that the start and end points are for the part of artery segments affected in $\mathbf{D}_{OCT}$ are obscure mostly in $\mathbf{D}_{MPR}$. Reference locations $R^i$ do not cover the start- and end-points of the entire segments. 

\begin{algorithm}[htb!]
\caption{Alignment between OCT and CCTA}\label{alg.align}
\KwIn{OCT image $X^i_{OCT}$, MPR image$X^i_{CCTA}$, reference locations $R^i$, OCT label $Y^i_{OCT}$}
\KwOut{MPR Label $Y^i_{MPR}$}
\BlankLine
\tcp{Step 1. Get base transform ratio via thickness from the images}
$ \gamma_{base} \gets \frac{GetSliceThickness(X^i_{OCT})}{GetSliceThickness(X^i_{MPR})}$\;
\tcp{Step 2. Calculate transform ratio via reference locations}
$ R^i \gets Sort(R^i) $\;
$ I^i_{OCT} \gets CalculateListIntervals(R^i[:,0]) $\;
$ I^i_{MPR} \gets CalculateListIntervals(R^i[:,1]) $\;
$ \Gamma \gets \frac{I^i_{MPR}}{I^i_{OCT}} $\;
\tcp{Step 3. Integrate transform ratio of head and tail}
\eIf{${len}(\Gamma) \geq 1$}
{
    $\Gamma$.insert($\Gamma$[0], 0) \;
    $\Gamma$.append($\Gamma$[-1]) \;
}{
    $\Gamma \gets [\gamma_{base}, \gamma_{base}]$ \;
}
\tcp{Step 4. Determine the segment's start- and end-point in MPR}
$ p_{start} = R^i[0, 1] - round(\Gamma[0] \cdot R^i[0, 0]) $\;
$ p_{end} = R^i[-1, 1] + round(\Gamma[-1] \cdot (len(X^i_{OCT}) - R^i[-1, 0])) $\;
$ R^i.insert((0, p_{start}), 0) $\;
$ R^i.append((len(X^i_{OCT}), p_{start})) $\;
\tcp{Step 5. Align annotions from OCT to MPR}
$ Y^i_{MPR} \gets [] $\;
\For{$(i = 0 \ to \ len(R^i) - 1)$}
{
    $ Y^i_{MPR} $.append(Merge($Y^i_{OCT}[R^i[i]:R^i[i+1]]$, $I^i_{MPR}[i]$))\;
}
return $ Y^i_{MPR}$\;
\end{algorithm}

\subsubsection{Alignment Process}
As described above, the alignment objects and reference locations are all set. For the convenience of introduction, we take a patient as an example. The alignment process $\mathcal{A}$ is essentially a spatial co-registration problem between $X^i_{OCT}$ and $X^i_{MPR}$. The interval of reference locations could provide enough spatial correspondence. Algorithm~\ref{alg.align} gives a detailed illustration of the alignment process $\mathcal{A}$, and the annotation $Y^i_{MPR}$ is generated. It must be noted that this method assumes that in each interval, the spatial mapping relationship is uniform. This introduces little deviations, but is acceptable.

\begin{figure}[t!]
\includegraphics[width=\textwidth]{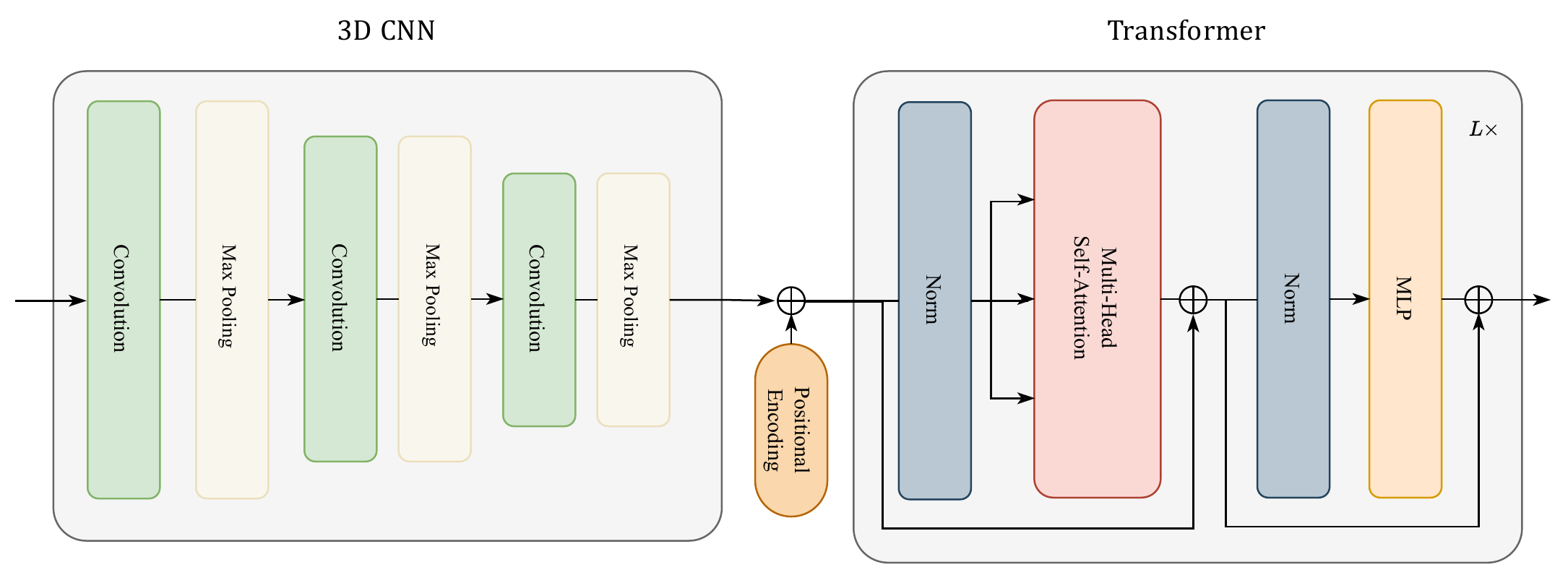}
\caption{The classification model's architecture consists of a shallow 3D CNN, and a Transformer follows. Sequences of MPR volumes are fed into it to predict lumen morphological characteristics and plaque composition.} 
\label{fig.networks}
\end{figure}

\subsection{Classification Model}
\label{sec.model}
After the alignment process, data $\mathbf{D}_{MPR}$ and substantial label $\mathbf{Y}_{MPR}$ get ready now for building a classification model. 
Previous work used a CNN network to extract image features and an RNN or Transformer to capture the relationships of volume sequence along the vessel centerline. The network ~\cite{ma2021transformer} combining CNN and Transformer achieves a new state-of-the-art in stenosis detection on MPR images. Inspired by ~\cite{ma2021transformer}, our model adopts the same network design ideas, shown in Fig~\ref{fig.networks}. It is committed to introducing more fine-grained knowledge, almost inaccessible for experts relying on CCTA scans alone.

The input of model $X^i_{MPR}$ is cut into a sequence of 3D volumes, with a shape of $L^i \times N \times D \times D$. $L^i$ denotes the length of the sequence, varying among cases. $D$ is the size of 2D slices, and $N$ is the number of slices in a 3D volume. Correspondingly, $Y^i_{MPR}$ has a size of $L^i \times C$, which has $C$ categories. 

As shown in Fig.~\ref{fig.networks}, a 3D-CNN $\phi (\cdot)$ is employed to extract local characteristics at a specific location in a coronary artery. These features of each 3D volume are fed into a Transformer $\sigma (\cdot)$, analyzing the dependence among the features from different positions. Then, the objective $\mathcal{L}$ of the model turns out to be:
\begin{equation}
    \mathcal{L} = - \sum_{j=1}^{C} {y_j \log(\sigma(\phi(X^i_{MPR})))}, \quad y_j \in Y^i_{MPR}
\end{equation}

\section{Experiment} 
\subsection{Implement Details}
Experiments are conducted on a dataset consisting of 55 CCTA scans and OCT images. With the efforts of two experts, OCT images are labeled healthy, calcified, lipid-rich, fibrous, thrombus, and stent. The annotation clarifies the lumen morphological characteristics and plaque composition slice-by-slice. 
As mentioned in Sec.~\ref{sec.model}, the extracted MPR images are split into sequences of 3D volumes, and we set its shape to $12\times21\times21$. All models are evaluated with five-fold cross-validation, and trained for 2000 epochs in each fold. Mini-batches of 8 sequences are used to minimize the loss function with Adam optimizer~\cite{kingma2014adam}, and a cosine annealing learning rate scheduler ~\cite{smith2017cyclical} initialized by \num{1e-6}. 

\begin{table}[t!]
    \centering
    \resizebox{\textwidth}{!}{ 
    \begin{tabular}{l | p{1.5cm} | p{1.5cm}<{\centering} p{1.5cm}<{\centering} p{1.5cm}<{\centering} p{1.5cm}<{\centering} p{1.5cm}<{\centering} p{1.5cm}<{\centering}}
    \toprule
    \multicolumn{2}{l}{\multirow{2}{*}{}} & \multicolumn{6}{c}{OCT} \\
    \cmidrule{3-8}
    \multicolumn{2}{c}{~} & Healthy & Calcified & Lipid-rich & Fibrous & Thrombus & Stent\\
    \midrule
    \multirow{3}{*}{CCTA} &  Non-cal. & \checkmark &  & \checkmark & \checkmark & \checkmark & \\
                          &  Calcified  &  & \checkmark & & & & \\
                          &  Stent      &  &  & & & & \checkmark\\
    \bottomrule
    \end{tabular}
    \caption{Correspondence of manual annotations of lumen morphological characteristics and plaque composition via CCTA and OCT.}
    \label{tab.label_consistency}
    }
\end{table}

\subsection{Verification of Annotation Differences}
The foundation of the O2CTA problem we propose for the first time is the distinct annotation differences between CCTA and OCT. Before presenting our model experimental results, we would like to quantify this difference first. Two experienced doctors carefully labeled the MPR images into non-calcified, calcified, and stent, providing lumen morphological characteristics and plaque composition from the perspective of CCTA. 

The categories correspond to those in OCT annotations, and the detailed relation are shown in Table~\ref{tab.label_consistency}. According to it, the consistency between CCTA and OCT annotations is evaluated on 1158 3D volumes. The result are shown in Table~\ref{Tab.confusion}. After calculation, the Kappa coefficient is 0.113, indicating a poor consistency. The large gap brings considerable improvement room, and this is where clinical value resides in our work.

\begin{table*}[t!]
\begin{floatrow}
\resizebox{\textwidth}{!}{ 
\capbtabbox{
    \begin{tabular}{l | p{1.3cm} | p{1.3cm}<{\centering} p{1.3cm}<{\centering} |p{1.3cm}<{\centering}}
    \toprule
    \multicolumn{2}{l}{\multirow{2}{*}{}} & \multicolumn{3}{c}{OCT}\\
    \cmidrule{3-5}
    \multicolumn{2}{c}{~}  & Positive & Negative & Sum\\
    \midrule
    \multirow{3}{*}{CCTA} &  Positive & 121 & 190 & 311 \\
                          &  Negative & 140 & 128 & 268 \\
                          \cmidrule{2-5}
                          &  Sum      & 261 & 318 & 1158 \\
    \bottomrule
    \end{tabular}
}{
 \caption{Consistency between manual annotations of lumen morphological characteristics and plaque composition via CCTA and OCT.}
 \label{Tab.confusion}
}
\capbtabbox{
    \begin{tabular}{l p{1.8cm}<{\centering} p{1.2cm}<{\centering} p{1.2cm}<{\centering}}
    \toprule
    Model & Volume Size & AUC & ACC  \\
    \midrule
    \multirow{3}{*}{3D CNN + Transformer} &  $6\times21\times21$ & 80.24 & 85.58  \\
                          &  $9\times21\times21$ & 80.73 & \pmb{86.57}  \\
                          &  $12\times21\times21$    & \pmb{83.06} & 86.20 \\
    \bottomrule
    \end{tabular}
}{
 \caption{Ablation study on the size of 3D volumes, which are the input of networks.}
 \label{Tab.ablation}
 \small
}
}
\end{floatrow}
\end{table*}

\subsection{Experimental Results}
\begin{figure}[t!]
\includegraphics[width=\textwidth]{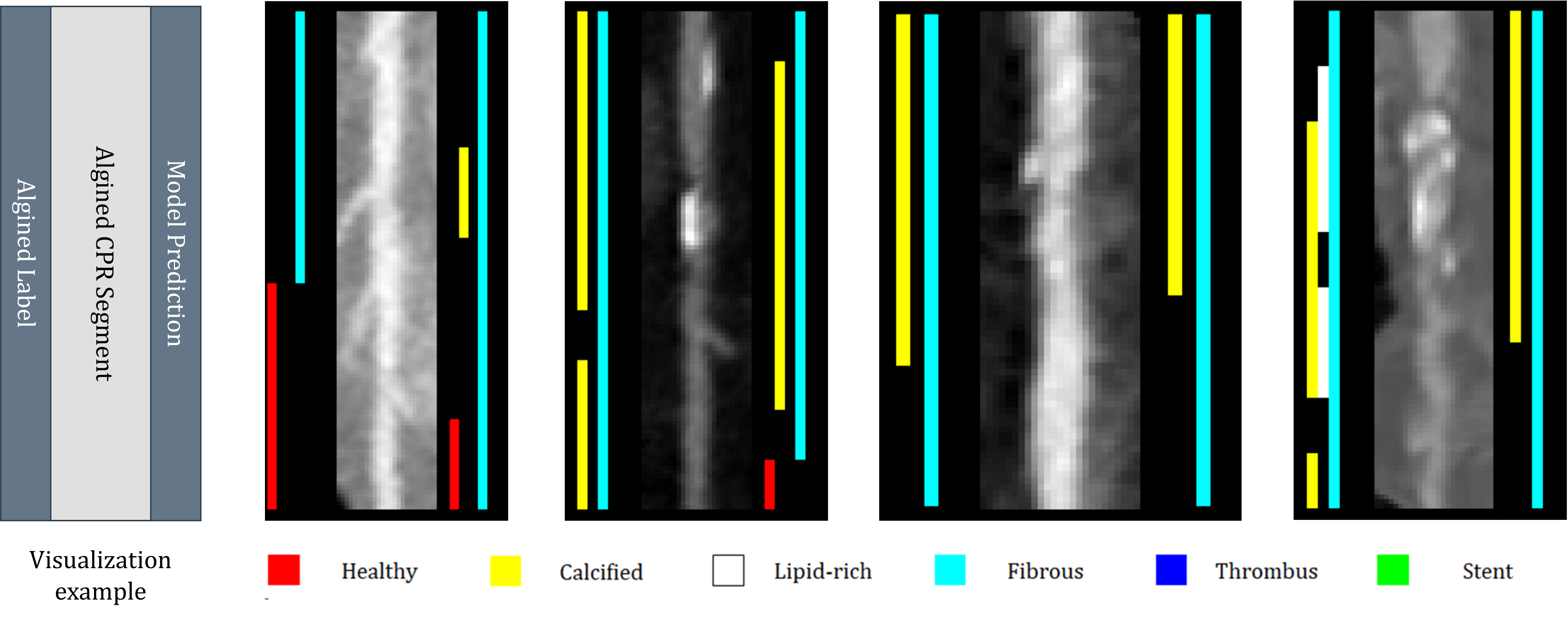}
\caption{Visualization of aligned label and classification results. The schematic plot on the left is the visualization example, where aligned MPR is placed in the middle, and aligned annotation and classification are set in the left and right, respectively.} 
\label{fig.visualize}
\end{figure}

To assess the effectiveness of the proposed solution for plaque lumen morphological characteristics and plaque composition classification, we evaluated the area under the ROC curve (AUC) and accuracy (ACC) based on classification results.  As shown in Table~\ref{tab.result}, 3D CNN analyzes segments independently. 3D RCNN connects a 3D CNN with an LSTM in series to processing a sequential MPR volumes. And our network achieved the best performances in terms of AUC and ACC metrics for different plaques. The qualitative visualization of plaque assessment result are shown in Fig.~\ref{fig.visualize}. 

\begin{table}[t!]
\caption{Evaluation results for lumen morphological characteristics and plaque composition classification.}
\label{tab.result}
\centering
\resizebox{\textwidth}{!}{ 
\begin{tabular}{l p{1.5cm}<{\centering} p{1.5cm}<{\centering} p{1.5cm}<{\centering} p{1.5cm}<{\centering} p{1.5cm}<{\centering} p{1.5cm}<{\centering} p{1.5cm}<{\centering} p{1.5cm}<{\centering}}
\toprule
\multirow{2}{*}{Models} & \multicolumn{7}{c}{AUC} & \multirow{2}{*}{ACC}\\
\cmidrule{2-8}
  & Healthy & Calcified & Lipid-rich & Fibrous & Thrombus & Stent & Mean & \\
\midrule
3D CNN &  76.27 & 74.53 & 56 & 71.88 & 84.23 & 65.71 & 71.44 & 84.34 \\
3D RCNN &  74.08 & 72.85 & 73.56 & 65.38 & 90.21 & 74.07 & 75.03 & 80.89 \\
3D CNN + Transformer &  \pmb{80.81} & \pmb{77.33} & \pmb{79.22} & \pmb{76.63} & \pmb{94.39} & \pmb{89.95} & \pmb{83.06} & \pmb{86.20} \\
\bottomrule
\end{tabular}
}
\end{table}
As shown in Table~\ref{Tab.ablation}, an ablation study is performed on the size of 3D volumes. The MPR images are split into volumes with a size of $6\times21\times21$, $9\times21\times21$, and $12\times21\times21$. And the results demonstrate 12 is the best choice of volume thickness. We hence use this choice in our experiments.

\section{Discussions}
Based on observations of differences in the imaging capabilities of CCTA and OCT, we infer and demonstrate that it would affect the coronary plaque analysis task, and raise the O2CTA problem further. The key to solving the O2CTA problems lies in annotation alignment and the design of classification neural networks. In this study, we introduce an alignment strategy based on reference locations and a network arranging 3D CNN and Transformer in sequence, which is experimentally demonstrated.

In annotation alignment, the current strategy assumes the spatial mapping relationship is uniform in each interval. More dedicated alignment methods, e.g., co-registration based on lumen morphology, can diminish or eliminate introduced deviations. Besides, the classification performance can benefit from strengthening the representation and learning ability of the model. From the perspective of future work, there is considerable room for improvement in the O2CTA problem.

%
%
%
%

\bibliographystyle{splncs04}
\bibliography{refe}

\end{document}